\documentstyle[aps,prd]{revtex}

\begin{document}

\draft

\tightenlines

\preprint{quantum-ph/0202xxx}

\title{Decoherence of Quantum Damped Oscillators}

\author{Sang Pyo Kim\footnote{Email address: sangkim@kunsan.ac.kr}}
\address{Department of Physics, Kunsan National University,
Kunsan 573-701, Korea}
\author{Ademir E. Santana\footnote{Email address:
santana@fis.ufba.br}}
\address{Instituto de Fisica, Universidade
Federal da Bahia, Campus de Ondina, 40210-340, Salvador, Bahia,
Brazil}
\author{F. C. Khanna\footnote{Email address:
khanna@phys.ualberta.ca}}
\address{Theoretical Physics Institute, Department of
Physics, University of Alberta, Edmonton, Alberta, Canada T6G
2J1\\ TRIUMF, 4004 Wesbrook Mall, Vancouver, British Columbia,
Canada, V6T 2A3}

\date{\today}

\maketitle

\begin{abstract}
Quantum dissipation is studied within two model oscillators, the
Caldirola-Kanai (CK) oscillator as an open system with one degree
of freedom and the Bateman-Feshbach-Tikochinsky (BFT) oscillator
as a closed system with two degrees of freedom. Though these
oscillators describe the same classical damped motion, the CK
oscillator retains the quantum coherence, whereas the damped
subsystem of the BFT oscillator exhibits both quantum decoherence
and classical correlation. Furthermore the amplified subsystem of the BFT
oscillator shows the same degree of quantum decoherence and
classical correlation.
\end{abstract}


\section{Introduction}

Physical systems are believed without doubt to obey quantum theory
that has been tested in various areas. Microscopic systems are
described most accurately by the quantum theory, whereas most of
the macroscopic systems are described also by classical theory to any
desired precision. The most prominent feature of quantum theory is
quantum coherence, the interference among superposed states. Then
the question may be raised: how quantum states of a system can have
the characteristic features of classical theory such as quantum
decoherence (loss of quantum coherence) and classical correlation
along classical trajectory? This quantum-to-classical transition
or classicality of quantum system has been an important problem
since the advent of quantum theory. Though not completely settled,
it is well known that the interaction of a quantum system with an
environment can lead to classicality of quantum decoherence and
classical correlation \cite{zeh,zur,joo,joo2,cald,unr,hu,han}. In
particular, dissipation is known to result in quantum decoherence
\cite{cald,unr,hu}. In most of literature, the quantum decoherence has
been studied for a system coupled to an environment or thermal
bath with many degrees of freedom.

In this paper we study quantum damped oscillators as a simple
model for the dissipative system. There are different Hamiltonian
representations for this damped oscillator. One representation is
the Caldirola-Kanai (CK) oscillator, which is a one-dimensional
system with an exponentially increasing mass \cite{cal,kan}. This
oscillator is an open system because its parameters such as mass
or frequency depend explicitly on time. The other representation
is the Bateman or Feshbach-Tikochinsky (BFT) oscillator, which
consists of a damped oscillator and an amplified oscillator
\cite{bat,mor,fes}. The second oscillator is a closed system as
the total energy is conserved and the energy dissipated from the
damped oscillator is transferred to the amplified one. These quantum
damped oscillators have been studied intensively as a model to
understand dissipation in quantum theory and the connection
between them has been found \cite{dek}. The CK oscillator has also
been investigated to find the characteristic features of quantum
states \cite{dod,col,lea,cer} (for review and references, see Ref.
\cite{um}). The BFT oscillator has also been used to study quantum
dissipation \cite{fes,dek,vit,vit2,vit3,san}.

The main purpose of this paper is to study the classicality of the
damped CK and BFT oscillators as a quantum mechanical system with
dissipation and few degrees of freedom. More concretely we
investigate quantum decoherence, the necessary condition, and
classical correlation, the sufficient condition for classicality.
For that purpose we first find the density matrix of the Gaussian
state for the CK oscillator and the reduced density matrix for the
damped part of the BFT oscillator and then apply the criterion on
measuring quantitatively quantum decoherence and classical
correlation. The measure of quantum decoherence is defined as the
ratio of the diagonal element to the off-diagonal element of the
(reduced) density matrix \cite{mori}. It is found that the density
matrix for the CK oscillator shows no quantum decoherence, whereas
the reduced density matrix for the BFT oscillator shows both
quantum decoherence and classical correlation for the damped
and amplified parts.

The organization of this paper is as follows. In Sec. II we find
the quantum state and density matrix of CK oscillator and
calculate the measure of quantum decoherence and classical
correlation. In Sec. III we find the density matrix and thereby
the reduced density matrix of the BFT oscillator. The measure of
decoherence is evaluated. In Sec. IV we study the amplified
oscillator, the opposite case of damped oscillator and investigate
the condition for classicality.

\section{CK Oscillator}

The CK oscillator is an an open system with the variable mass
$m(t) = me^{\gamma t/m}$ \cite{cal,kan}
\begin{equation}
H_{\rm CK} = \frac{1}{2 me^{\gamma t/m}} p_x^2 + \frac{m\omega^2
e^{\gamma t/m}}{2}x^2, \label{ck ham}
\end{equation}
and has the classical equation of motion for dissipative system
\begin{equation}
\ddot{x} + \frac{\gamma}{m} \dot{x} + \omega^2 x = 0. \label{dam
eq}
\end{equation}
It is assumed that the quantum theory of the damped oscillator is
prescribed by the time-dependent Schr\"{o}dinger equation
\begin{equation}
i \hbar \frac{\partial}{\partial t} \Psi(x, t) = \hat{H}_{\rm CK}
(t) \Psi(x, t). \label{sch eq}
\end{equation}
The wave functions are found in various methods
\cite{dod,col,lea,cer} (for review and references, see Ref.
\cite{um}). Lewis and Riesenfeld introduced an invariant operator
for the general time-dependent oscillator, whose eigenstate is an
exact quantum state up to a time-dependent phase factor
\cite{lew}.

Following Refs. \cite{kim,kim2,kim3}, we introduce a pair of first
order operators in position and momentum
\begin{eqnarray}
\hat{a} (t) &=&  i [ u^* (t) \hat{p}_x - \dot{u}^* (t)\hat{x} ],
\nonumber\\ \hat{a}^{\dagger} (t) &=& - i [ u (t) \hat{p}_x -
\dot{u} (t) \hat{x} ], \label{an-cr}
\end{eqnarray}
and require them to satisfy the quantum Liouville-von Neumann
equation
\begin{eqnarray}
i \hbar \frac{\partial}{\partial t} \hat{a} (t) + [  \hat{a} (t) ,
\hat{H}_{\rm CK} (t)] = 0, \nonumber\\ i \hbar
\frac{\partial}{\partial t} \hat{a}^{\dagger} (t) + [
\hat{a}^{\dagger} (t) , \hat{H}_{\rm CK} (t)] = 0. \label{ln eq}
\end{eqnarray}
Then $u$ satisfies the classical equation of motion (\ref{dam
eq}). The Wronskian condition
\begin{equation}
\hbar m e^{\gamma t/m} \Bigl[\dot{u}^* (t) u (t) - \dot{u} (t) u^*
(t) \Bigr] = i, \label{wron}
\end{equation}
guarantees the standard commutation relation for all times
\begin{equation}
[\hat{a} (t), \hat{a}^{\dagger} (t)] = 1.
\end{equation}
We are interested in the underdamped motion given by
\begin{equation}
u(t) = \frac{1}{\sqrt{2 \hbar m  \Omega}} e^{- \gamma t/(2m)} e^{
- i\Omega t},
\end{equation}
where
\begin{equation}
\Omega = \sqrt{\omega^2 - \Bigl(\frac{\gamma}{2m} \Bigr)^2}, \quad
(\gamma \leq 2 m \omega) . \label{omega}
\end{equation}

The number operator defined by
\begin{equation}
\hat{N} (t) = \hat{a}^{\dagger} (t) \hat{a} (t)
\end{equation}
also satisfies Eq. (\ref{ln eq}) and yields the number state as an
exact quantum state
\begin{equation}
\hat{N} (t) \vert n, t \rangle = n \vert n, t \rangle.
\end{equation}
The wave function for the number state that satisfies Eq.
(\ref{sch eq}) is given by \cite{kim3}
\begin{equation}
\Psi_{n} (x, t) = \Bigl(\frac{m \Omega e^{\gamma t/m}}{\pi \hbar}
\Bigr)^{1/4} \frac{e^{- i (n+ 1/2) \Omega t}}{\sqrt{2^{n} n!}}
H_{n} \Bigl(\sqrt{\frac{m \Omega e^{\gamma t/m}}{\hbar}}  x \Bigr)
\exp \Bigl[- e^{\gamma t/m} \Bigl(\frac{m \Omega}{2 \hbar} + i
\frac{\gamma}{4 \hbar} \Bigr) x^2 \Bigr], \label{har wav}
\end{equation}
where $H_{n}$ is the Hermite polynomial. The wave function has the
dispersion relations
\begin{eqnarray}
\langle \hat{x}^2 \rangle &=& \hbar^2 u^* (t) u (t) =
\frac{\hbar}{2 m \Omega } e^{- \gamma t/m}, \nonumber\\ \langle
\hat{p}_x^2 \rangle &=& \hbar^2 m^2(t) \dot{u}^* \dot{u} =
\frac{\hbar m \omega^2}{2 \Omega} e^{\gamma t/m}. \label{dis}
\end{eqnarray}
The more the system dissipates in time, the more the wave function
becomes sharply peaked around $x = 0$, whereas the wave function
is more dispersed in momentum space. The uncertainty is a constant
\begin{equation}
(\Delta x) (\Delta P_x) = \frac{\hbar \omega}{2 \Omega} \geq
\frac{\hbar}{2}, \label{unc}
\end{equation}
and the Hamiltonian expectation value is also a constant.
\begin{equation}
\langle 0, t \vert \hat{H}_{\rm CK} (t) \vert 0, t \rangle =
\frac{\hbar \omega^2}{\Omega} \geq \frac{\hbar \omega}{2}.
\label{ham ex}
\end{equation}

The density matrix of the Gaussian wave function (\ref{har wav}),
the ground state with $n = 0$, has the form
\begin{eqnarray}
\rho_{\rm CK} (x', x, t) &=& \Psi_0 (x') \Psi_0^*(x) \nonumber\\
&=& \Bigl(\frac{m \Omega e^{\gamma t/m}}{\pi \hbar}\Bigr)^{1/2}
\exp \Bigl[ - \Gamma_c x_c^2 - \Gamma_{\delta} x_{\delta}^2 -
\Gamma_{\mu} x_c x_{\delta} \Bigr], \label{ck den}
\end{eqnarray}
where
\begin{equation}
x_c = \frac{1}{2} (x' + x), \quad x_{\delta} = \frac{1}{2} (x' -
x),
\end{equation}
and
\begin{equation}
\Gamma_c = \Gamma_{\delta} = \frac{m \Omega}{\hbar} e^{\gamma
t/m}, \quad \Gamma_{\mu} = - i \frac{\gamma}{\hbar} e^{\gamma
t/m}.
\end{equation}
The off-diagonal element, the coefficient of $x_{\delta}^2$,
measures the degree of quantum coherence, {\it i.e.}, the interference
between two different trajectories. The representation-independent
measure of quantum decoherence \cite{mori} is now given by
\begin{equation}
\delta_{\rm QD} = \frac{1}{2}
\sqrt{\frac{\Gamma_c}{\Gamma_{\delta}}} = \frac{1}{2}. \label{dec}
\end{equation}
The condition for quantum decoherence $(\delta_{\rm QD} \ll 1)$
shows no decoherence for the CK oscillator. Likewise, the measure
of classical correlation
\begin{equation}
\delta_{\rm CC} = \sqrt{\frac{\Gamma^2_c
\Gamma^2_{\delta}}{\Gamma_{\mu}^* \Gamma_{\mu}}} = \frac{(m
\Omega)^2}{\hbar \gamma} e^{\gamma t /m}, \label{cor}
\end{equation}
shows no classical correlation conditioned by $\delta_{\rm CC} \ll
1$, except for the case of $\Omega \approx 0$ in the large-damping
limit $\gamma \approx 2 m \omega$. Even in this case the
exponentially growing factor dominates at later times and
classical correlation is lost. The zero-damping limit $(\gamma =
0)$ of a pure harmonic oscillator does not lead to any classical
correlation with the infinite $\delta_{\rm CC}$ as expected. The
CK oscillator does achieve neither quantum decoherence nor
classical correlation.

In summary, the CK oscillator does not have the genuine properties
of dissipative systems, though its equation of motion does show
such a damping effect. First, the energy defined by the
expectation value of the Hamiltonian operator does not have any
damping factor that implies the dissipation of energy. Second, the
uncertainty does not grow as the evolution proceeds. Third, there
is neither quantum decoherence nor classical correlation
regardless of the magnitude of damping factor.

\section{BFT Oscillator}

The one-dimensional CK oscillator has a constant expectation value
of the Hamiltonian. To be a genuine dissipative system, the energy
of the damped subsystem of the system must be dissipated away and
transferred to another subsystem. This means that the damped
oscillator may be properly described by a two-dimensional system,
one subsystem of which dissipates the energy and the transferred
energy amplifies the other subsystem. Such a model has been
suggested long ago by Bateman \cite{bat} and later 
by Feshbach and Tikochinsky \cite{mor,fes}. 
The BFT oscillator is described by the Lagrangian
\begin{equation}
L = m \dot{x} \dot{y} + \frac{\gamma}{2} ( x \dot{y} - \dot{x} y)
- k xy. \label{lag}
\end{equation}
The one subsystem with $x$ variable obeys the damped equation of
motion (\ref{dam eq}) where $\omega = \sqrt{k/m}$. The other
subsystem obeys the equation for an amplified oscillator
\begin{equation}
\ddot{y} - \frac{\gamma}{m} \dot{y} + \omega^2 y = 0. \label{en
eq}
\end{equation}
The energy of $y$ increases as $e^{\gamma t/m}$. The Hamiltonian
is given by
\begin{equation}
H_{\gamma} = \frac{1}{m} p_x p_y + \frac{\gamma}{2m} (y p_y - x
p_x) + \Omega^2 xy, \label{ft ham}
\end{equation}
where $p_x = m \dot{y}$ and $p_y = m \dot{x}$.

In the limit of zero-dissipation $(\gamma = 0)$, the BFT
oscillator is the sum of two decoupled oscillators with opposite
signs
\begin{equation}
H_{0} = \frac{1}{2m} p_{\xi}^2 + \frac{k}{2} \xi^2 - \frac{1}{2m}
p_{\zeta}^2 - \frac{k}{2} \zeta^2,
\end{equation}
where
\begin{equation}
\xi = \frac{1}{\sqrt{2}} (x + y), \quad \zeta =
\frac{1}{\sqrt{2}}(- x + y).
\end{equation}
The ground state of each oscillator leads to the zero energy and
a density matrix
\begin{eqnarray}
\rho (x', y', x, y) &=& \Bigl(\frac{m \omega}{\pi \hbar} \Bigr)
\exp \Bigl[- \frac{m \omega}{2 \hbar} (\xi'^{2} + \zeta'^{2} +
\xi^2 + \zeta^2) \Bigr] \nonumber\\&=& \Bigl(\frac{m \omega}{\pi
\hbar} \Bigr) \exp \Bigl[- \frac{m \omega}{2 \hbar} (x'^{2} +
y'^{2} + x^2 + y^2) \Bigr]. \label{zero den}
\end{eqnarray}

For the dissipation case $(\gamma \neq 0)$, the density matrix
satisfies the quantum Liouville-von Neumann equation, whose
coordinate representation is given by
\begin{eqnarray}
i \hbar \frac{\partial t}{\partial t} \rho (x', y', x, y) = \Bigl[
- \frac{\hbar^2}{m} \Bigl(\frac{\partial^2}{\partial x' \partial
y'} - \frac{\partial^2}{\partial x \partial y} \Bigr) + i
\frac{\hbar \gamma }{2m} \Bigl(x' \frac{\partial}{\partial x'} -
y' \frac{\partial}{\partial y'} + x \frac{\partial}{\partial x} -
y \frac{\partial}{\partial y} \Bigr) \nonumber\\ + m \Omega^2
(x'y' - xy) \Bigr] \rho (x', y', x, y). \label{den eq}
\end{eqnarray}
The density matrix quadratic in $x', y', x, y$ has the general
form
\begin{eqnarray}
\rho (x', y', x, y) = N \exp \Bigl[ - e^{\gamma t/m} (A^* x'^{2} +
A_1 x' x + A x^2) - e^{- \gamma t/m} (B^* y'^{2} + B_1 y' y +
By^2) \nonumber\\ - C (x'y' + xy) - (D x'y + D^* xy') \Bigr],
\label{den}
\end{eqnarray}
where the coefficients satisfy the set of equations
\begin{eqnarray}
\dot{A} &=& i \frac{\hbar}{m} (2 A C - A_1 D^*), \nonumber\\
\dot{B} &=& i \frac{\hbar}{m} (2 B C - B_1 D), \nonumber\\
\dot{A}_1 &=& i \frac{2 \hbar}{m} (A D - A^* D^*), \nonumber\\
\dot{B}_1 &=& i \frac{2 \hbar}{m} (B D^* - B^* D), \nonumber\\
\dot{C} &=& i \frac{\hbar}{m} (4 A B  + C^2 - A_1 B_1 - D^* D) - i
\frac{m \Omega^2}{\hbar}, \nonumber\\ \dot{D} &=& i
\frac{2\hbar}{m} (A_1 B - A^* B_1). \label{eqs}
\end{eqnarray}
There is a symmetry under $x \leftrightarrow y$ and $\gamma
\leftrightarrow - \gamma$, which is in fact the time-reversal
symmetry of the Hamiltonian (\ref{ft ham}). Hence we can set $A =
B^*$ and $A_1 = B_1^*$. Also in the zero-dissipation limit, by
comparing the density matrices (\ref{zero den}) and (\ref{den}) we
find that $A_1, C, D$ approach zero. A particular solution is
found to be
\begin{eqnarray}
A (\gamma) &=& B^*(\gamma) = \Bigl[\Bigl(\frac{m \Omega}{2 \hbar}
\Bigr)^2 + \frac{D^*(\gamma) D(\gamma)}{4} \Bigr]^{1/2} e^{i (\pi
-  \theta)}, \nonumber\\ D (\gamma) &=& |D (\gamma)| e^{i \theta},
\nonumber\\ A_1 &=& B_1 = C = 0, \label{part sol}
\end{eqnarray}
where
\begin{equation}
D (\gamma \rightarrow 0) \rightarrow 0.
\end{equation}

Now the density matrix from the particular solution (\ref{part
sol}) becomes
\begin{equation}
\rho (x', y', x, y) = N \exp \Bigl[ - e^{\gamma t/m} (A^* x'^{2} +
A x^2 ) - e^{- \gamma t/m} (A y'^{2} + A^* y^2) - \Bigl(D x'y +
D^* xy' \Bigr) \Bigr], \label{den2}
\end{equation}
where
\begin{equation}
N = \frac{1}{\pi} \Bigl[ (A + A^*)^2 - \frac{1}{4} (D + D^*)^2
\Bigr]^{1/2}.
\end{equation}
Letting $y' = y$ and integrating over $y$, one obtains the reduced
density matrix for the damped subsystem of $x'$ and $x$:
\begin{equation}
\rho_{\rm red} (x', x) = N_1  \exp \Bigl[ - e^{\gamma t/m}
\Bigl\{(A^* x'^{2} + A x^2) - \frac{(Dx' + D^* x)^2 }{4 (A + A^*)}
\Bigr\} \Bigr], \label{red den}
\end{equation}
where
\begin{equation}
N_1 = N \times \sqrt{\frac{\pi e^{\gamma t/m}}{(A + A^*)}}.
\end{equation}
Similarly, the reduced density matrix for the amplified subsystem is
obtained by using the symmetry $x \rightarrow y$ and $\gamma
\rightarrow - \gamma$:
\begin{equation}
\rho_{\rm red} (y', y) = N_2 \exp \Bigl[ - e^{-\gamma t/m} \Bigl\{
(A y'^{2} + A^* y^2) - \frac{(D y + D^* y')^2 }{4 (A + A^*)}
\Bigr\} \Bigr], \label{red den amp}
\end{equation}
where
\begin{equation}
N_2 = N \times \sqrt{\frac{\pi e^{- \gamma t/m}}{(A + A^*)}}
\end{equation}

Then the reduced density matrix (\ref{red den}) is written in the
form
\begin{equation}
\rho_{\rm red} (x', x) = N_1 \exp \Bigl[ - \Gamma_c x_c^2 -
\Gamma_{\delta} x_{\delta}^2 - \Gamma_{\mu} x_c x_{\delta} \Bigr],
\label{red den2}
\end{equation}
where
\begin{eqnarray}
\Gamma_c (|D|, \theta) &=&  e^{\gamma t /m} \Bigl[(A + A^*) -
\frac{(D + D^*)^2}{ 4 (A + A^*)} \Bigr], \nonumber\\
 \Gamma_{\delta} (|D|, \theta) &=& e^{\gamma t /m} \Bigl[(A + A^*) - \frac{(D -
D^*)^2}{ 4 (A + A^*)} \Bigr], \nonumber\\ \Gamma_{\mu} (|D|,
\theta) &=& e^{\gamma t /m} (-2) \Bigl[(A - A^*) + \frac{D^2 -
D^{*2}}{ 4 (A + A^*)} \Bigr]. \label{gamma}
\end{eqnarray}
As the density matrix (\ref{den2}) and the reduced one (\ref{red
den}) depend only two real parameters $|D|$ and $\theta$,  the
measure of quantum decoherence is given by
\begin{eqnarray}
\delta_{\rm QD} &=& \frac{1}{2} \sqrt{\frac{\Gamma_c (|D|,
\theta)}{\Gamma_{\delta} (|D|, \theta)}} \nonumber\\ &=&
\frac{1}{2} \sqrt{\frac{(m\Omega/\hbar)^2}{(m\Omega/\hbar)^2
\cos^2 \theta + |D|^2}}. \label{dec2}
\end{eqnarray}
and that of classical correlation by
\begin{eqnarray}
\delta_{\rm CC} &=& \sqrt{\frac{\Gamma^2_c (|D|, \theta)
\Gamma^2_{\delta} (|D|, \theta) }{\Gamma_{\mu}^* (|D|, \theta)
\Gamma_{\mu} (|D|, \theta)}} \nonumber\\ &=& \frac{1}{2} |\cot
\theta| \Bigl[\Bigl(\frac{m \Omega}{\hbar}\Bigr)^2 \cos^2 \theta +
|D|^2 \Bigr]. \label{cor2}
\end{eqnarray}
If the condition
\begin{equation}
\theta \approx \frac{\pi}{2}, \quad |D| > \frac{m \Omega}{\hbar}
\label{dec con}
\end{equation}
is satisfied, the density matrices (\ref{den2}) and (\ref{red den}) achieve a
significant degree of quantum decoherence $\delta_{\rm QD} < 1/2$
and a sufficient degree of classical correlation $\delta_{\rm CC}
\ll 1$. This is the case of the large dissipation.

In summary, the BFT oscillator has the density matrix which
achieves quantum decoherence as well as almost complete classical
correlation. This means that the BFT oscillator may be the quantum
analog of a classical dissipative oscillator.

\section{Amplified Oscillator}

We now consider the opposite case of the damped oscillator, that
is, the amplified oscillator, which is physically motivated by an
unstable system, for instance, the second order phase transition
during the spinodal instability. Quantum decoherence has not been
observed for an unstable, exponentially growing, single oscillator \cite{kim4}.
This oscillator has the Hamiltonian
\begin{equation}
H_y = \frac{1}{2 me^{- \gamma t/m}} p_y^2 + \frac{m\omega^2 e^{-
\gamma t/m}}{2}y^2.
\end{equation}
The time-dependent annihilation and creation operators are given
by
\begin{eqnarray}
\hat{b} (t) &=&  i [ v^* (t) \hat{p}_y - \dot{v}^* (t)\hat{y} ],
\nonumber\\ \hat{b}^{\dagger} (t) &=& - i [ v (t) \hat{p}_y -
\dot{v} (t) \hat{y} ],
\end{eqnarray}
where
\begin{equation}
\ddot{v} - \frac{\gamma}{m} \dot{v} + \omega^2 v = 0.
\end{equation}
Then the exponentially growing solution is given by
\begin{equation}
v(t) = \frac{1}{\sqrt{2 \hbar m  \Omega_y}} e^{ \gamma t/(2m)} e^{
- i\Omega t},
\end{equation}
where $\Omega$ is given by Eq. (\ref{omega}). The harmonic
oscillator wave functions are obtained from Eq. (\ref{har wav}) by
replacing $\gamma$ by $(- \gamma)$:
\begin{equation}
\Psi_{n} (y, t) = \Bigl(\frac{m \Omega e^{- \gamma t/m}}{\pi
\hbar} \Bigr)^{1/4} \frac{e^{- i (n+ 1/2) \Omega t}}{\sqrt{2^{n}
n!}} H_{n} \Bigl(\sqrt{\frac{m \Omega e^{- \gamma t/m}}{\hbar}}  y
\Bigr) \exp \Bigl[- e^{- \gamma t/m} \Bigl(\frac{m \Omega}{2
\hbar} - i \frac{\gamma}{4 \hbar} \Bigr) y^2 \Bigr]. \label{gr har
wav}
\end{equation}

Repeating the steps in Sec. II,  the dispersion relations are
given by
\begin{eqnarray}
\langle \hat{y}^2 \rangle &=& \frac{\hbar}{2 m \Omega_y } e^{
\gamma t/m}, \nonumber\\ \langle \hat{p}_y^2 \rangle &=&
\frac{\hbar m \omega^2}{2 \Omega_y} e^{- \gamma t/m}.
\end{eqnarray}
So the uncertainty relation and the Hamiltonian expectation value
have the same values as Eqs. (\ref{unc}) and (\ref{ham ex}),
respectively. As the ground state wave function $(n = 0)$ in Eq.
(\ref{gr har wav}) is obtained by replacing $\gamma$ by $ (-
\gamma)$ and all the steps are the same in Sec. II, we find the
measures for quantum decoherence and classical correlation
\begin{eqnarray}
\delta_{\rm QD} &=& \frac{1}{2}, \nonumber\\ \delta_{\rm CC} &=&
 \frac{(m \Omega)^2}{\hbar \gamma} e^{- \gamma t
/m}.
\end{eqnarray}
The wave functions for the amplified oscillator are certainly
classically correlated as shown in Ref. \cite{kim4}. But this does
not mean that the wave functions achieve quantum decoherence
because $\delta_{\rm QD} = 1/2$.

Now we turn to the amplified subsystem of the BFT oscillator in Sec.
III. The reduced density matrix for the amplified subsystem can be
written in the form
\begin{equation}
\rho_{\rm red} (y', y) = N_2 \exp \Bigl[ - \Gamma^y_c y_c^2 -
\Gamma^y_{\delta} y_{\delta}^2 - \Gamma^y_{\mu} y_c y_{\delta}
\Bigr],
\end{equation}
where
\begin{eqnarray}
\Gamma^y_c (|D|, \theta) &=&  e^{- \gamma t /m} \Bigl[(A + A^*) -
\frac{(D + D^*)^2}{ 4 (A + A^*)} \Bigr], \nonumber\\
 \Gamma^y_{\delta} (|D|, \theta) &=& e^{- \gamma t /m} \Bigl[(A + A^*) - \frac{(D -
D^*)^2}{ 4 (A + A^*)} \Bigr], \nonumber\\ \Gamma^y_{\mu} (|D|,
\theta) &=& e^{- \gamma t /m} (2) \Bigl[(A - A^*) + \frac{D^2 -
D^{*2}}{ 4 (A + A^*)} \Bigr].
\end{eqnarray}
Therefore, for the particular solution (\ref{part sol}) the
measures for quantum decoherence and classical correlation for the
amplified subsystem are given by the same Eqs. (\ref{dec2}) and
(\ref{cor2}) for the damped subsystem. This may be expected from
the symmetry of the Hamiltonian under $x \leftrightarrow y$ and
$\gamma \leftrightarrow - \gamma$. There is also the symmetry of
reduced density matrices of $x$ and $y$ under $\gamma
\leftrightarrow - \gamma$ and $D ( - \gamma) = D^* (\gamma)$. When
the condition (\ref{dec con}) is satisfied, the amplified subsystem
also achieves both quantum decoherence and classical correlation.
Such classicality of wave functions has been observed for the
system of an unstable amplified oscillator coupled to a stable
oscillator in a quantum phase transition model \cite{kim4}. The
BFT oscillator provides an exactly solvable model for both quantum
decoherence and classical correlation and may shed some light in
understanding how the classicality of quantum systems can be
achieved.

\section{Discussion}

We have studied the CK  and BFT oscillators as the quantum analogs
of a classical dissipative oscillator that obeys the classical
damped equation of motion. The CK oscillator is a one-dimensional
oscillator with an exponentially increasing mass in time. The wave
function of the CK oscillator is sharply peaked around the origin
with the exponentially decreasing position-dispersion. However,
the momentum-dispersion increases exponentially so that the
uncertainty of both position and momentum is constant and has the
value greater than the minimum value of Heisenberg uncertainty
relation. Similarly, the Hamiltonian expectation value has a
constant value greater than the minimum value of a harmonic
oscillator without the damping factor. Further, CK oscillator does
not achieve both quantum decoherence and classical correlation.

On the other hand, the BFT oscillator is a two-dimensional system,
one subsystem describing the damped oscillator and the other
describing an amplified oscillator. In this oscillator
the energy dissipated away from the damped oscillator is
transferred to the other so that the total energy is conserved. We
have found two parameter-dependent density matrix (\ref{den2}),
whose reduced density matrix for the damped oscillator or the
amplified oscillator shows not only quantum decoherence but also
classical correlation. The quantum decoherence of the BFT
oscillator may be understood as a consequence of the coupling
between the damped and amplified modes.

The fact that the interaction or coupling of the system is
indispensable for quantum decoherence has been observed in the
two-oscillator model with the Hamiltonian \cite{han}
\begin{equation}
H = \frac{1}{2m} (p_1^2 + p_2^2) + \frac{m}{2} (\omega_1^2 x_1^2 +
\omega_2^2 x_2^2) + \lambda x_1 x_2.
\end{equation}
The reduced density matrix for $x_1$ is given by
\begin{equation}
\rho_{\rm red} (x'_1, x_1) = \Bigl(\frac{1}{\pi D} \Bigr)^{1/2}
\exp \Bigl[- \frac{1}{D}(x_c^2 + \Gamma_{\delta} x_{\delta}^2)
\Bigr],
\end{equation}
where
\begin{eqnarray}
D &=& \cosh \eta - \sinh \eta \cos (2 \vartheta), \nonumber\\
\Gamma_{\delta} &=& \cosh^2 \eta - \sinh^2 \eta \cos^2 (2
\vartheta),
\end{eqnarray}
where
\begin{eqnarray}
e^{\eta} &=& \frac{m (\omega_1^2 + \omega_2^2) +
\sqrt{m^2(\omega_1^2 - \omega_2^2)^2 + 4 \lambda^2}}{2 \sqrt{m^2
\omega_1^2 \omega_2^2 - \lambda^2}}, \nonumber\\ \tan (2
\vartheta) &=& \frac{2 \lambda}{m (\omega_2^2 - \omega_1^2)}.
\end{eqnarray}
The measure of decoherence
\begin{equation}
\delta_{\rm QD} = \frac{1}{2 \sqrt{\cosh^2 \eta - \sinh^2 \eta
\cos^2 (2 \vartheta)}}
\end{equation}
has a value $\delta_{\rm QD} \leq 1/2$, the equality corresponding to the
zero mode-mixing, and has small values $\delta_{\rm QD} \ll 1/2)$
for large mixing angles $(\vartheta \approx \pi/4$ and $\eta \gg
1)$. However, the measure of classical correlation is infinite,
the same as the quantum state of a simple harmonic oscillator.
Therefore we may conclude that a system with dissipation (damping)
interacting with another system can achieve both quantum
decoherence and classical correlation.

\acknowledgments

S.P.K. and A.E.S. would like to appreciate the warm hospitality of
the Theoretical Physics Institute, Univ. of Alberta. S.P.K. also
would like to thank D.N. Page for helpful discussions on quantum
decoherence and Prof. Y.S. Kim for useful information. F.C.K. was
supported by NSERC of Canada, S.P.K. by KRF under Grant No.
2000-015-DP0080, and A.E.S. by CNPq of Brazil.

\end{document}